\documentclass[fleqn,10pt]{wlscirep}
\usepackage[utf8]{inputenc}
\usepackage[T1]{fontenc}
\usepackage{tabularx}
\usepackage{booktabs}
\title{Dynamic Caustics by Ultrasonically Modulated Liquid Surface}

\author[1]{Koki Nagakura}
\author[2,3,*]{Tatsuki Fushimi}
\author[1]{Ayaka Tsutsui}
\author[2,4]{Yoichi Ochiai}

\affil[1]{Graduate School of Comprehensive Human Sciences, University of Tsukuba, Tsukuba, 305-8550, Japan}
\affil[2]{Institute of Library, Information and Media Science, University of Tsukuba, Kasuga Campus Kasuga 1-2, Tsukuba, 305-8550, Ibaraki, Japan}%
\affil[3]{R\&D Center for Digital Nature, University of Tsukuba, Kasuga Campus Kasuga 1-2, Tsukuba, 305-8550, Ibaraki, Japan}
\affil[4]{Pixie Dust Technologies, Inc, Tokyo, 104-0028, Japan}%

\affil[*]{tfushimi@slis.tsukuba.ac.jp} 


\begin{abstract}
This paper presents a method for generating dynamic caustic patterns by utilising dual-optimised holographic fields with Phased Array Transducer (PAT). Building on previous research in static caustic optimisation and ultrasonic manipulation, this approach employs computational techniques to dynamically shape fluid surfaces, thereby creating controllable and real-time caustic images. The system employs a Digital Twin framework, which enables iterative feedback and refinement, thereby improving the accuracy and quality of the caustic patterns produced.
This paper extends the foundational work in caustic generation by integrating liquid surfaces as refractive media. This concept has previously been explored in simulations but not fully realised in practical applications. The utilisation of ultrasound to directly manipulate these surfaces enables the generation of dynamic caustics with a high degree of flexibility. The Digital Twin approach further enhances this process by allowing for precise adjustments and optimisation based on real-time feedback.
Experimental results demonstrate the technique's capacity to generate continuous animations and complex caustic patterns at high frequencies. Although there are limitations in contrast and resolution compared to solid-surface methods, this approach offers advantages in terms of real-time adaptability and scalability. This technique has the potential to be applied in a number of areas, including interactive displays, artistic installations and educational tools.
This research builds upon the work of previous researchers in the fields of caustics optimisation, ultrasonic manipulation, and computational displays. Future research will concentrate on enhancing the resolution and intricacy of the generated patterns.
\end{abstract}

\begin{document}

\flushbottom
\maketitle
%
%
\thispagestyle{empty}

\noindent Please note: Abbreviations should be introduced at the first mention in the main text – no abbreviations lists. The suggested structure of the main text (not enforced) is provided below.


\section*{Introduction}
\begin{figure}[ht]
  \centering
  \includegraphics[width=\textwidth]{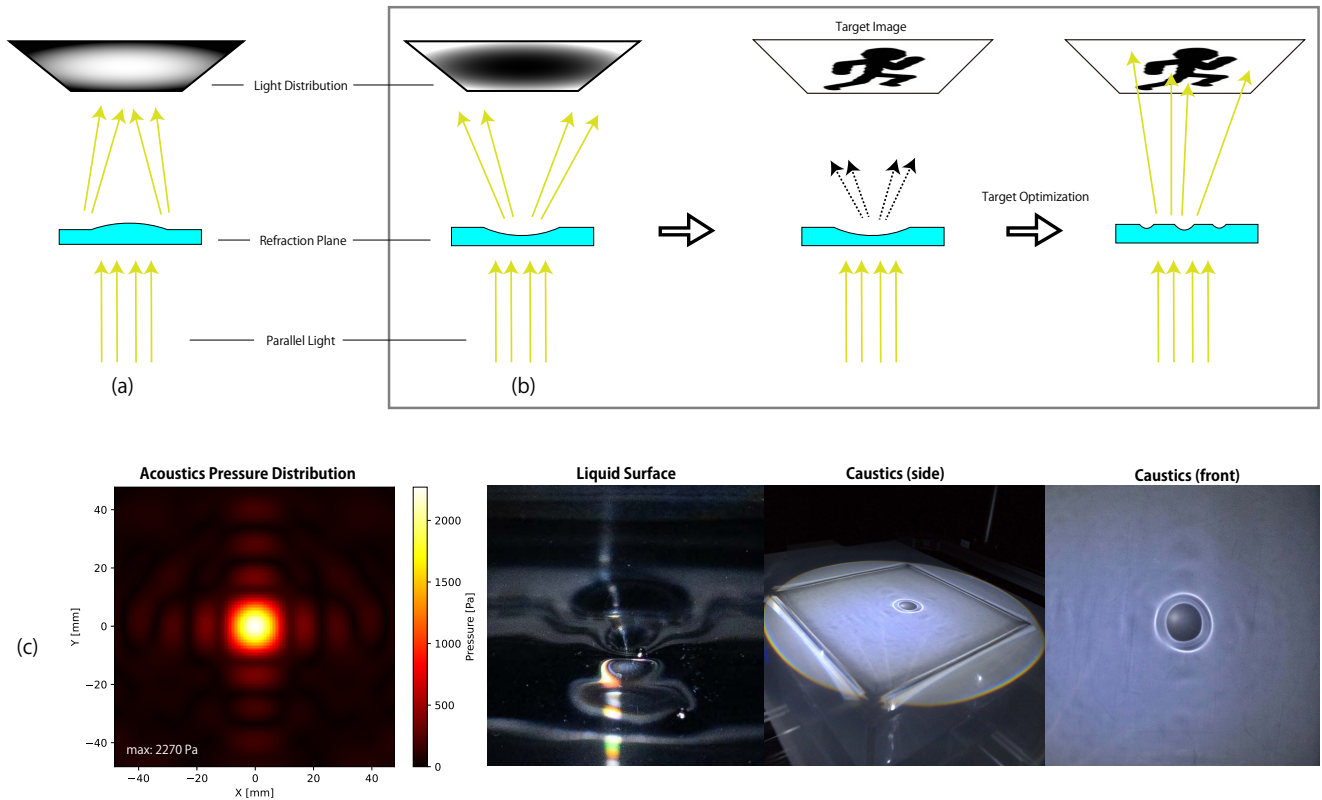}
  \caption{The core principle of forming caustic patterns through acoustic modulation of a liquid surface. (a, b) illustrate how the liquid surface deforms into convex or concave shapes due to acoustic pressure, resulting in caustic pattern formation. When parallel light rays pass through the deformed refractive surface, they are either focused inward (convex) or dispersed outward (concave), creating distinct caustic patterns. This iterative optimization process adjusts the liquid surface shape to align the refracted light distribution with a desired target image. The results displayed in (c) are visualizations of a numerically simulated process. This process calculates the phase delay of each transducer in the acoustic phased array based on the distance to the focal point, thereby estimating the liquid surface shape and generating the caustic pattern. The leftmost image shows the acoustic pressure distribution used to deform the liquid surface, followed by the corresponding liquid surface deformation. Finally, the resulting caustic pattern is shown from the side and front views, revealing a bright ring with a darker center that closely matches the intended target image.}
  \label{mechanism} 
\end{figure}

In optics, caustics is defined as ``the envelope of light rays that have been reflected or refracted by a curved surface or object, or the projection of that envelope of rays on another surface'' \cite{Weinstein1969-de}. When light enters an object with a high refractive index from the air, its direction of travel is altered in accordance to Snell's law. A convex surface focuses light towards the center, thereby creating a bright spot in the vicinity of the focal point and a region of reduced brightness at the edges (Figure \ref{mechanism} (a)). In contrast, concave surfaces bend light outward, brightening the edges and darkening the center (Figure \ref{mechanism} (b)).

Researchers have investigated methods for computing the shape of a light-refracting surface in order to generate specific projected images utilising caustics. For instance, methods have been proposed to calculate smooth refractive surfaces capable of producing desired image using differential geometry approaches \cite{Yue2012-nn, Yue2014-hs}. Schwartzburg et al. proposed a method to calculate refractive surfaces with smooth regions, infinite strength singularities, and wholly black regions by solving the optimal transport problem \cite{Schwartzburg2014-oy}. However, these approaches often involve static solid surfaces acrylic, limiting their ability to produce dynamic animations. Additionally, they require milling processes on materials to achieve caustics in real-world applications. 
While Yue et al.~\cite{Yue2014-hs} optimized continuous and smooth refractive surfaces using the Poisson equation, Schwartzburg et al.~\cite{Schwartzburg2014-oy} introduced piecewise smooth surfaces, resulting in high-contrast caustic images with stark black regions and optical density singularities. Our method, due to the use of a liquid surface, aligns more closely with the approach by Yue et al., leading to more gradual transitions in the refractive surface and, consequently, lower contrast.

Previous studies have proposed the use of liquid surfaces as refractive media for caustic generation, however, such work has been limited to in-silico simulations and has not been fully realized in practical applications \cite{Suzuki2019-xj}. Here, we experimentally demonstrate dynamic caustics capable of depicting text and images by directly shaping fluid surfaces using ultrasound, achieving dynamic caustics in practise where previous studies were limited to simulations. To achieve this, we utilized a Phased Array Transducer (PAT), a technology commonly used for non-contact tactile displays \cite{Hoshi2010, Long2014, monnai2014haptomime}, volumetric displays\cite{Hirayama2019, Fushimi2019a} and digital microfluidics \cite{Koroyasu2023-ws}. 

Our approach to generating dynamic caustic patterns relies on the precise manipulation of fluid surfaces using the PAT. The PAT used in this study is equipped with 256 transducers arranged in a 16 $\times$ 16 matrix, which allows for the generation of complex acoustic pressure fields \cite{Morales2021-wy}. These pressure fields are capable of deforming the liquid surface in a controlled manner, thereby influencing the refraction of light and producing caustic patterns. As shown in Figure 2, The PAT is positioned 200 mm above the liquid surface, which is composed of 1 kg of transparent silicone oil, contained within a transparent acrylic tank. 

Shin-Etsu Chemical KF-96H-10000 silicone oil (cosmetic name: Dimethicone) was selected for its high viscosity of 10,000 mm²/s at 25°C, which suppresses bubble formation under high acoustic pressures and stabilizes liquid surface deformation. Its specific gravity is 0.975, and its refractive index ranges from 1.399 to 1.403 (measured with respect to the sodium D line), making it suitable for precise light refraction. Additionally, it has a speed of sound of approximately 987 m/s and a surface tension of 20–21 mN/m.

The experiment was conducted at approximately 25℃, where the physical properties of silicone oil vary with temperature. Specifically, the specific gravity of the KF-96H series decreases from 0.975 at 25℃ to 0.960 at 50℃, while the viscosity of KF-96H-10000 drops from 10,000 mm\textsuperscript{2}/s at 25℃ to 9,800 mm\textsuperscript{2}/s at 30℃.

The silicone oil's high viscosity suppresses fluid flow, reducing unintended fine vibrations and surface waves, stabilizing the caustic pattern generation. However, excessive viscosity may hinder the liquid surface's ability to quickly respond to ultrasonic pressure changes, potentially reducing responsiveness.

\begin{figure}[ht]
  \centering
  \includegraphics[width=\textwidth]{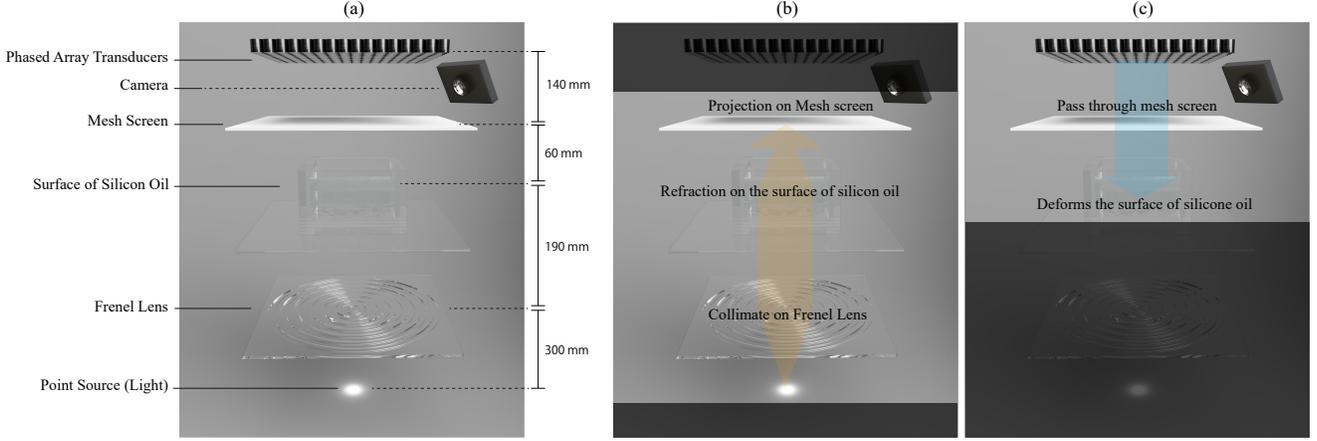}
  \caption{Schematic representation of the experimental setup and the principles behind caustic generation. (a) Overall system configuration including the phased array transducer, camera, mesh screen, silicone oil tank, Fresnel lens, and point light source. (b) Illustration of how light from below is collimated by the Fresnel lens and then refracted at the silicone oil surface, projecting a caustic pattern onto the mesh screen. (c) Depiction of how ultrasonic waves emitted from the phased array transducer pass through the mesh screen, deforming the silicone oil surface and dynamically modulating the formed caustic pattern.}
  \label{explanation} 
\end{figure}

\begin{figure}[ht]
  \centering
  \includegraphics[width=\textwidth]{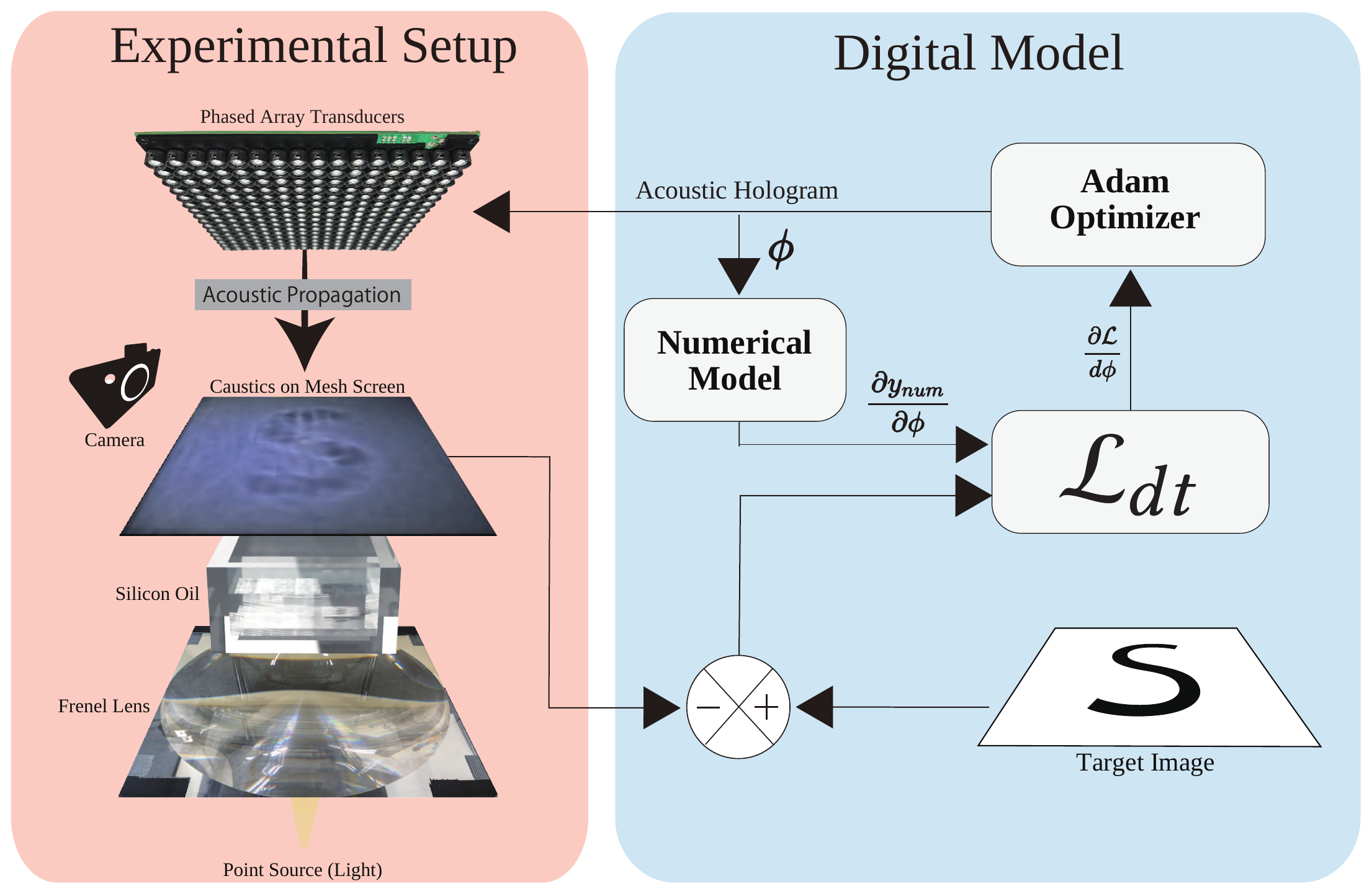}
  \caption{Schematic diagram of the experimental setup and in-situ caustics optimization using a Digital Twin. The resulting caustics are captured by a camera at an angle to the screen and used for feedback. The captured caustics are compared to the target image, with the difference fed into the loss function, and the derivatives of the numerical model are calculated using automatic differentiation for real-time optimization.}
  \label{overview} 
\end{figure}

The system setup includes a vertically arranged configuration in which the PAT is directed downward towards the liquid medium (Figure \ref{explanation}). A parallel light source is created using a point LED light source, which is then collimated by a Fresnel lens to produce uniform illumination from below the tank. As the ultrasound waves emitted by the PAT interact with the liquid surface, they generate acoustic pressure fields that deform the surface, leading to the formation of caustic patterns when the light is refracted by these deformations. The acoustic pressure induces concave deformations on the liquid surface. As shown in Figure \ref{mechanism} (b), these concave surfaces bend light outward, generating shadows. In Figure \ref{mechanism} (c), a single focal point of acoustic pressure deforms the surface, resulting in a caustic pattern with a dark center, surrounded by light focused outward. 


\subsection*{Numerical Optimization to Caustics Image} \label{sec:Calculation Method of Sound Pressure Distribution}
To project the desired target image as caustics, an efficient numerical optimization scheme is desirable. Through trial and error, we observed that the acoustic pressure field approximates the resultant caustics field well and, therefore, opted to optimize the caustics image via acoustic hologram optimization. Recent progress in acoustic hologram optimization led to the ability to generate more complex acoustic fields than a single focus point, and we employ Diff-PAT \cite{acoustic_fushimi_2021}, as it was recently demonstrated that it can efficiently use experimental feedback to improve acoustic hologram accuracy\cite{Fushimi2024-ih}.

To optimize the acoustic pressure distribution based on the target image; we begin by calculating the total sound pressure $p_t(x, y)$ at a point $(x, y)$ on the plane using:

\begin{equation}
\label{pressure}
p_t(x, y) = \sum_{m=0}^{M} \left[ \frac{P_{ref}}{d(x, y, x_t, y_t)} D(\theta) e^{i(kd(x, y, x_t, y_t)+\phi_m)} \right],
\end{equation}

where $M$ is the total number of transducers, and $P_{ref}$ represents the amplitude of the reference pressure. Furthermore, $d(x, y, x_t, y_t)$ denotes the Euclidean distance between the point $(x, y)$ on the plane and the position $(x_t, y_t)$ of the transducer, while $D(\theta) = \frac{2J_1(kr \sin(\theta))}{kr \sin(\theta)}$ is the far-field directivity function of the transducer, with $\theta$ being the angle between the transducer's normal and the point $(x, y)$. Here, $J_1$ is the Bessel function of the first kind of order 1, and $r=5$ mm is the transducer radius. Furthermore, $k$ is the wavenumber, given by $k = \frac{2\pi f}{c_0}$, where it is assumed that the speed of sound in the air ($c_0$) is 346 m / s. Lastly, $\phi_m$ represents the phase delay of the $m$-th transducer.

We then calculate the minimization loss function using the target image $P_{target}$ and $P_{pred}$ within the same Region of Interest (ROI). It is important to note that the target image should be input such that white is represented as 0 and black as 255. This is because, as illustrated in Figure \ref{mechanism}, caustics are generated such that the shadows, i.e., the black areas, approximate the acoustic pressure field. Therefore, when optimizing the acoustic pressure field from the target image, it is essential to ensure that the acoustic pressure field is generated in the black areas of the target image. Generally, Python libraries such as OpenCV represent white as 255 and black as 0, thus it may be necessary to invert the image to achieve the correct black and white representation in some cases.

\begin{equation}
L_{num} = \sum_{(x, y) \in S} \left| P_{target, n} - P_{pred, n} \right|.
\label{loss_function_digital}
\end{equation}

In the loss function (Equation \ref{loss_function_digital}), S represents the set of coordinates for each grid defined on the plane where the sound pressure calculation is performed. The subscript $n$, on which $P_{target, n}$, and $P_{pred, n}$, depend, represents the number of optimization steps by the Adam optimizer. $P_{target}$ and $P_{pred}$ are vectorized images. The normalized target pressure is given by

\begin{equation}
P_{target, n} = \frac{P_{target} - \min(P_{target})}{\max(P_{target}) - \min(P_{target})},
\end{equation}
while the normalized numerically simulated pressure field, based on equation \ref{pressure}, is defined as

\begin{equation}
P_{pred, n} = \frac{P_{pred} - \min(P_{pred})}{\max(P_{pred}) - \min(P_{pred})}.
\end{equation}

By defining the loss function in this way, $P_{pred}$ is appropriately normalized while the optimization is performed, balancing the accuracy of the maximum pressure and the pressure distribution. The acoustic hologram, $\phi_m$, is optimized using the Adam optimizer \cite{kingma2014adam}, an algorithm widely adopted for its efficient iterative updates in applications such as machine learning. The initial phase estimate for each transducer is generated randomly, and the gradient necessary for optimization is computed through reverse-mode automatic differentiation with respect to the objective function $L_{num}$.

The optimization runs for a fixed 1000 steps, without utilizing convergence criteria. This approach ensures that phase adjustments for the transducers form the desired sound pressure distribution over the two-dimensional plane within the predetermined step limit. The resulting pressure distribution closely replicates the target image, and the simulated loss demonstrates stable behavior after approximately 400 steps, as shown in Figure \ref{result_experiment} (Simulation Loss).

\subsection*{Time Averaging}
While the loss function has converged well and the generated acoustic pressure field closely resembles the target image, we observe that the generated caustics are sensitive to unevenness in the field, resulting in local bright spots. To mitigate this issue, we employ a time-averaging method by updating the PAT at a high frequency, which helps to even out these local bright spots. This approach aligns with recent methods proposed by Elizondo et al., who suggested similar techniques to enhance the spatial resolution of acoustic pressure fields in general \cite{elizondo2023}.

We calculated the time-averaged pressure amplitude using:
\begin{equation}
    P_{\text{avg}} = \frac{1}{F} \sum_{f=1}^{F} \| P_{pred} \|
\end{equation}
where $F$ is the number of frames. This time-averaged pressure field can be substituted into equation \ref{loss_function_digital} for this averaged pressure. The improved caustics images with the varying number of frames (3, 9, 24 frames) are shown in Figure~\ref{result_experiment} b, c, and d, respectively. The latest PAT can reach up to an update frequency of 10,000 Hz \cite{Plasencia2020}.

\subsection*{Enhancing Caustic Generation with Digital Twin}
In order to further enhance the quality of the generated caustics, we perform experimental optimization. The experimental optimization of acoustic holograms presents a challenge, given that typical PAT systems comprise hundreds of transducers, and obtaining finite differences in such systems is time-consuming and inefficient. However, it is possible to approximate the gradient of the loss function in an experiment by combining the derivative obtained via automatic differentiation with the experimentally obtained values, as demonstrated by Fushimi et al.~\cite{Fushimi2024-ih}. Since the derivative of the loss function is approximated using automatic differentiation, only a single image is required per step for the optimization, significantly increasing the speed and efficiency of the optimization process.

In this approach, the generated caustics are captured using a camera fixed between the PAT and the tank, as illustrated in Fig.~\ref{explanation}. Since the caustic image is captured from an angle, a perspective transform is employed in OpenCV. The transform matrix required was obtained experimentally using a checkerboard pattern. The warped image is then cropped and rotated to fit the size of the target image. Furthermore, to remove non-caustic artifacts from the image, we utilize a calibration image captured prior to the display of caustics. The resulting image, designated $C_{img}$, is then passed to the Digital Twin optimization scheme, as depicted in Fig.~\ref{overview}.

Rather than normalizing the pressure amplitude, we opted to utilize a cosine similarity loss function to compare the similarity between the target image and the experimentally obtained caustics \cite{deep_lee_2022}. As with numerical optimization, it is crucial to ensure that the target image is input such that white is represented as 0 and black as 255.

\begin{equation}
L_{dt} = 1 - \frac{\langle P_{dt}, P_{target} \rangle}{\| P_{dt} \| \| P_{target} \|}
\end{equation}

where $\langle P_{pred}, P_{target} \rangle$ denotes the inner product of $P_{pred}$ and $P_{target}$, and $ \|P_{pred} \| $ along with $\| P_{target} \| $ represent their respective norms. The Digital Twin uses the images taken to optimise the caustics directly, and the use of cosine similarity prevents the sound pressure from becoming too loud in some areas. 

The experimentally obtained images are input into the Digital Twin framework as follows. It is important to ensure that the experimentally obtained images are also input as the target images, with white set to 0 and black set to 255.

\begin{equation}
    P_{dt} = P_{num} + G(C_{img} - P_{num})
\end{equation}

where \(P_{num}\) represents the acoustic pressure obtained through numerical calculations, and \(C_{img}\) denotes the experimentally obtained caustics images.
\(G\) is used to prevent halt tracking of \(C_{img} - P_{num}\) by employing the \texttt{tf.stop\_gradient()} function in TensorFlow. This ensures that the term is treated as a constant and unaffected by automatic differentiation.

\begin{figure}[t]
\centering
\includegraphics[width=\textwidth]{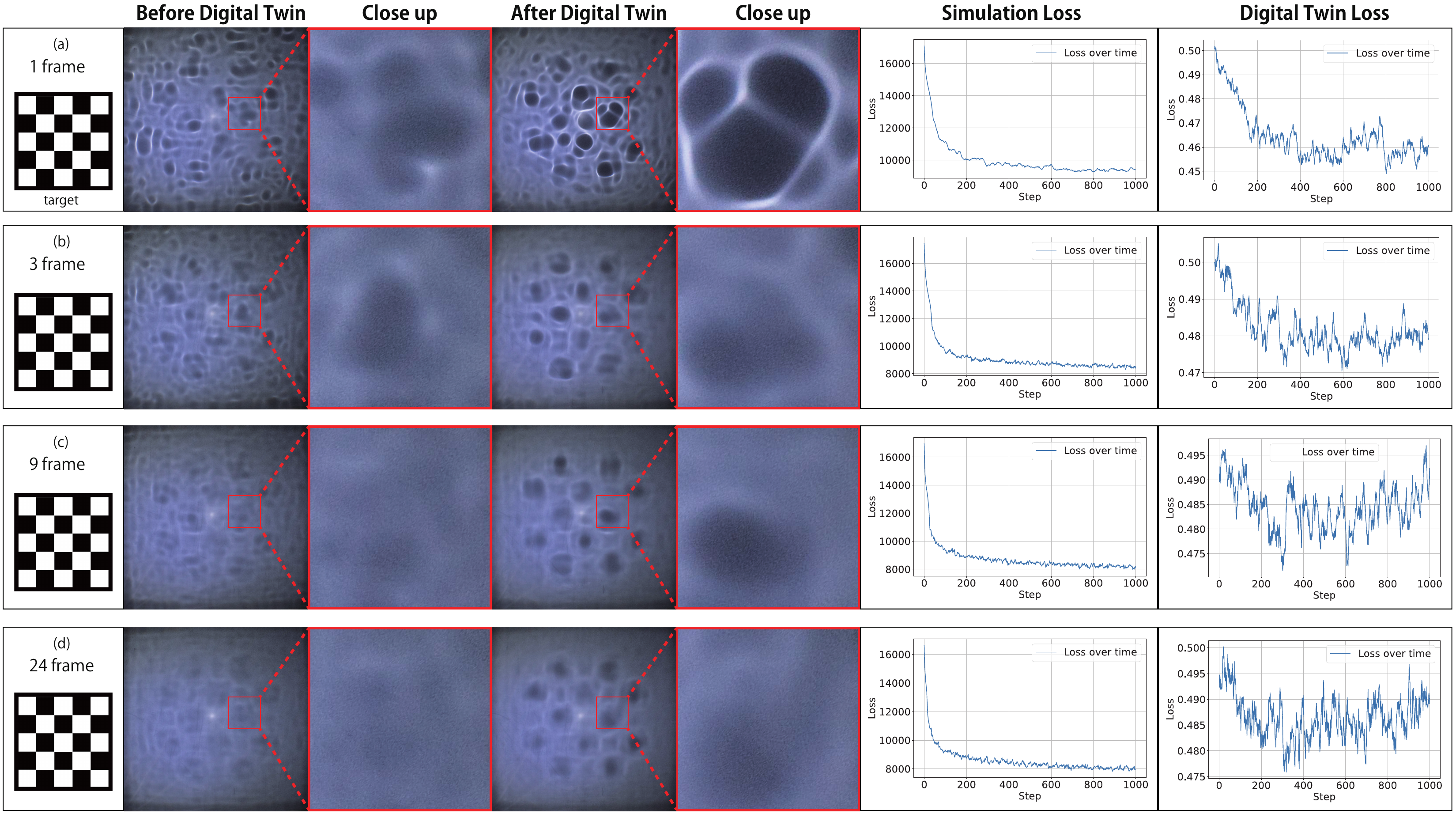}
\caption{Experimental results for the sound pressure distribution-based focusing technique with varying numbers of frames: 1, 3, 9, and 24. Each row displays, from left to right, the target pattern, visualization of the focus area before and after Digital Twin application (both overall and close-up views), and the loss function during optimization for both numerical simulation and the Digital Twin approach.}
\label{result_experiment} 
\end{figure}

The caustic images with Digital Twin optimisation are illustrated in Fig.~\ref{result_experiment}a-d. In contrast to the numerical simulation, the loss value does not converge as smoothly in the Digital Twin optimisation. However, the contrast between the white and black regions in the checkerboard is significantly enhanced with the application of Digital Twin.

\section*{Results and Discussion}
\subsection*{Animating Caustics}
\begin{figure}[htp]
  \centering
  \includegraphics[width=\textwidth]{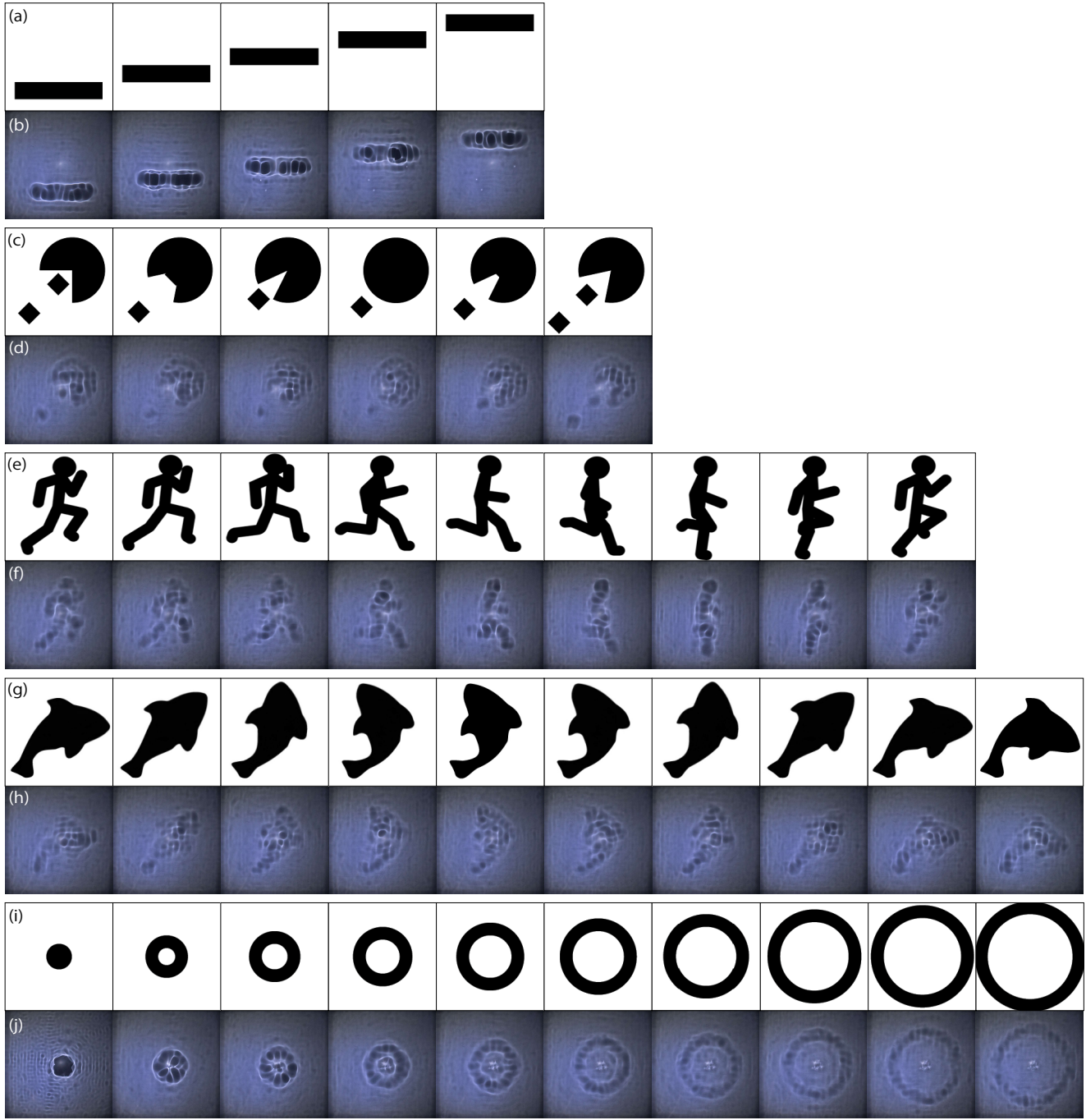}
  \caption{Comparison of target outcome used for animation and the generated caustics, presented frame by frame. The actual animation, created by projecting these frames sequentially, is included in the Supplemental material video. Each frame was generated by superimposing the sound pressure of 9 frames and then processed using the Digital Twin. (a)(b): Animation of parallel flowing sticks. (c)(d): Animation of a character preying on and sucking a square object. (e)(f): Animation of a running stick figure. (g)(h): Animation of a fish swimming and twisting its body. (i)(j): Animation showing ripples spreading from the center.}
  \label{animation}
\end{figure}

The prime advantage of this method is its ability to alter caustics temporally. In this instance, the caustics are presented in a sequence of individual frames, rather than in the conventional sequence of frames that would be expected in an animated sequence. The actual animation, which is generated by projecting the frames in succession, is included in the supplementary materials video. Each frame was created by superimposing the sound pressure of nine frames in order to optimize the initial sound pressure distribution, which was then processed using Digital Twin technology. The resulting caustic animations are presented in Figure~\ref{animation}. Animations a-b, c-d, and e-f depict parallel flowing sticks, a character preying on a square object while sucking it, a running stick figure, a fish swimming while twisting its body, and an animation showing ripples spreading out from the center, respectively. Given the dynamic nature of the caustic image, we strongly encourage readers to view the original video, which can be found in the supplementary material.

\subsection*{Contrast and Loss Behavior in Multi-Frame Processing}
While other quantitative metrics such as PSNR (Peak Signal-to-Noise Ratio) and SSIM (Structural Similarity Index) are commonly used for image quality assessment, we found them unsuitable for evaluating the current quality of our generated caustics. This is primarily because these metrics are designed to compare images with similar characteristics and are sensitive to pixel-level differences. However, in the current stage of our system, the generated caustic patterns, while visually recognizable, still exhibit significant differences in their textural properties compared to the target images. These differences stem from factors such as the inherent limitations of acoustic-based liquid surface deformation, the presence of noise as discussed previously, and the resolution limits of our current setup. Consequently, PSNR and SSIM tend to yield low scores that do not accurately reflect the perceived visual quality or the progress achieved in generating dynamic caustics. Therefore, we opted for Weber Contrast as a more pragmatic metric for this study, as it focuses on the relative luminance difference between the target and its background, which aligns better with the characteristics of our current caustic images \cite{Peli1990-ah}.

\begin{equation}
C = \frac{\Delta L}{L}
\end{equation}

Here, $\Delta L$ represents the change in luminance of the target relative to the uniform background luminance $L$. For our analysis, we divided the caustic image into a target area and a background area based on the target image, using the average brightness of each area for the calculation.

Table~\ref{result_contrast} shows improvements after applying Digital Twin processing across all frame counts. For instance, in the case of frame 1, the contrast increased by 231\% from 0.039 to 0.129. However, as the number of frames increased, a slight reduction in contrast was observed. This can be attributed to the averaging effects of acoustic pressure fields over multiple frames, which smooth the resulting caustics but simultaneously suppress fine features and sharp edges.

Despite the contrast improvements, the loss function during optimization displayed a notable behavior, particularly when increasing the number of frames. As shown in Figure~\ref{result_experiment} Digital Twin Loss, the loss initially decreases and stabilizes around 200-400 steps but begins to rise again during later stages. This trend becomes more pronounced for higher frame counts, such as 9 and 24 frames. The differences between the numerical optimization process and the Digital Twin optimization likely contribute to this phenomenon.

In numerical optimization, the target is a simple acoustic pressure distribution, and the loss function minimizes the absolute error between the predicted and target pressure fields. Conversely, the Digital Twin optimization uses experimentally captured caustic patterns as the target and employs cosine similarity as the loss function. The two approaches differ not only in their respective loss formulations but also in the nature of the target images. While numerical optimization operates purely in the acoustic domain, the Digital Twin incorporates real-world factors such as light refraction, surface tension, viscosity, and experimental noise. These physical complexities make it more challenging for the loss to converge smoothly, particularly in multi-frame scenarios where the acoustic field becomes increasingly intricate.

Another contributing factor is the optimization algorithm itself. The Adam optimizer, used in both numerical and Digital Twin optimizations, updates parameters based on momentum and adaptive learning rates. This can cause the loss to oscillate near local minima, especially in multi-frame processing where the complexity of the acoustic pressure distribution and its interactions with the liquid surface are amplified. Additionally, time-averaging the pressure fields over multiple frames further increases the deviation between the predicted and target caustics, as finer details in the experiment may not be preserved accurately.

\begin{figure*}[ht]
  \includegraphics[width=\textwidth]{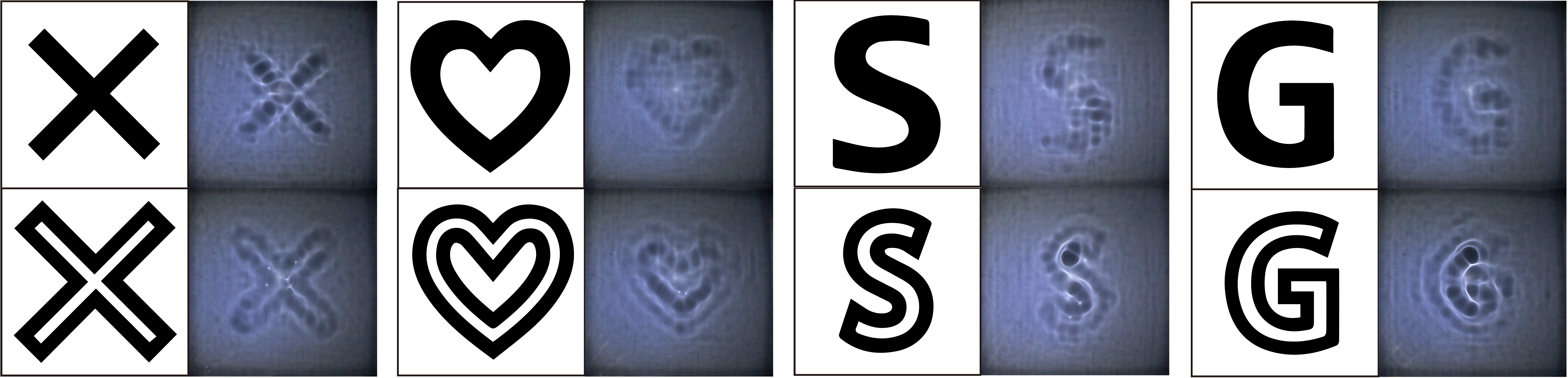}
  \caption{Caustic generation results. In all cases, the left image is the target image, and the right image shows the experimentally generated caustic pattern.}
  \label{Result_Focus_and_shadow}
\end{figure*}

Although the loss function does not exhibit smooth convergence in Digital Twin optimization, the contrast between the white and black regions of the checkerboard target is significantly improved, as shown in Figure~\ref{Result_Focus_and_shadow}. This suggests that the loss function, while useful for optimization, does not fully capture the perceptual quality of the caustics. Further refinement of the loss function to balance optical alignment and acoustic pressure accuracy may help address these issues. Improvements in the modeling of physical effects such as viscosity and light refraction could also contribute to more stable optimization results.

\begin{table}[h]
\centering
\caption{Weber Contrast}
\label{result_contrast}
\begin{tabular}{ccc}
\toprule
\textbf{Frame} & \textbf{Before Digital Twin} & \textbf{After Digital Twin} \\ \midrule
1  & 0.039 & 0.129 \\
3  & 0.049 & 0.093 \\
9  & 0.030 & 0.094 \\
24 & 0.026 & 0.080 \\ \bottomrule
\end{tabular}
\end{table}

\subsection*{Resolution}
\begin{figure*}[ht]
  \centering
  \includegraphics[width=\textwidth]{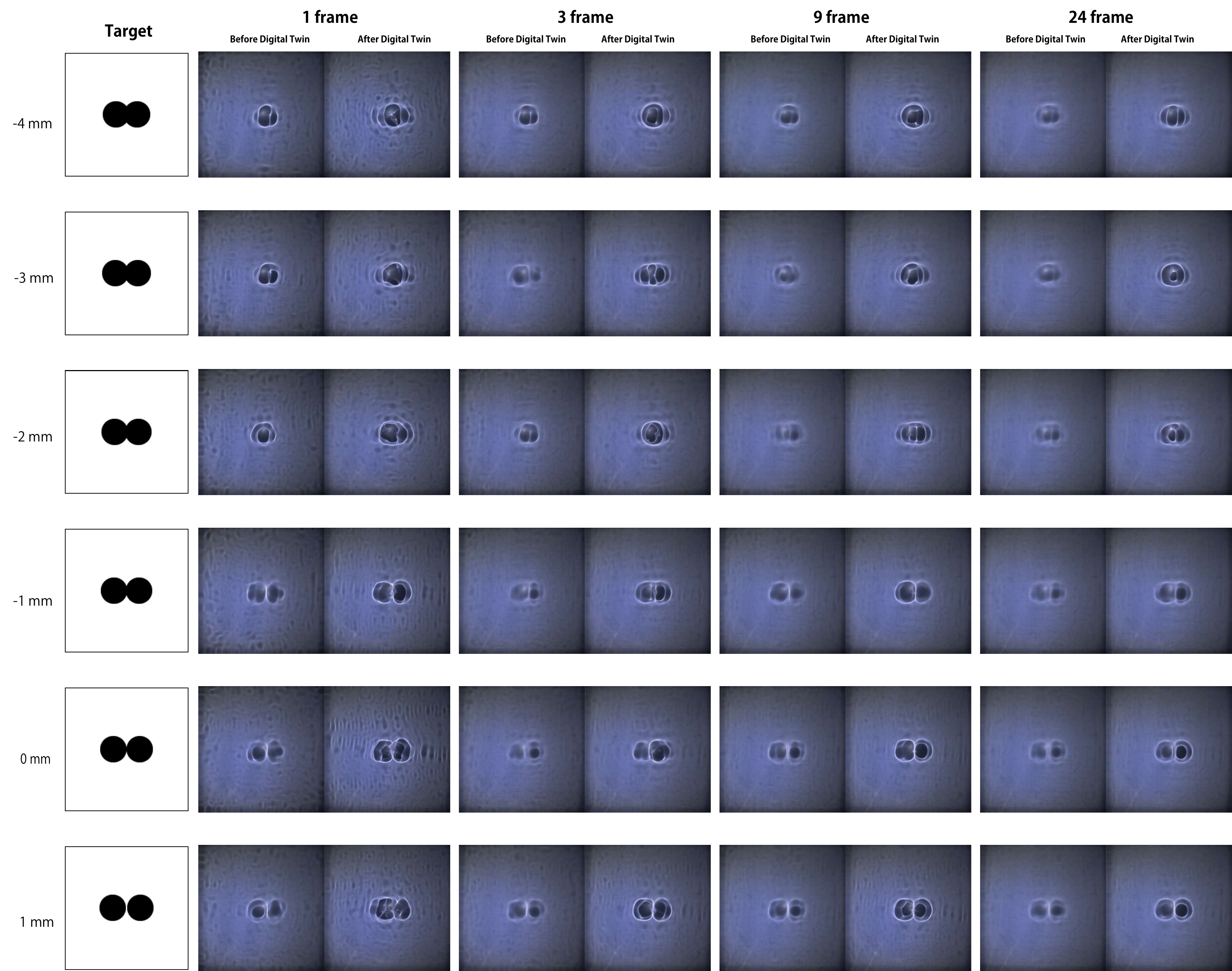}
  \caption{Resolution comparison using two circles with a radius of 10 mm, generated simultaneously with varying center-to-center distances. The distances range from 1 mm (bottom row) to -4 mm (top row, indicating overlap), increasing by 1 mm per row. The second row from the bottom shows the circles touching at their outer edges. Images were captured with 1, 3, 9, and 24 frames, both before and after applying the Digital Twin process.}
  \label{result_twopoint}
\end{figure*}

Our experiments focused on evaluating the resolution of caustic patterns generated through optimized sound pressure distributions. Figure \ref{result_twopoint} presents the results of varying the distance betweenda two circles to assess their resolution.

We generated two circles, each with a radius of 10 mm, and varied the distance between them from 1 mm to $-4$ mm (where they increasingly overlap). Images were captured at different frame counts (1, 3, 9, 24 frames) both before and after applying Digital Twin processing. The findings reveal that when the regions are of an optimized size, they can be distinctly resolved, whether adjacent or separated. Even with an overlap as small as 1 mm, the regions maintain their shapes after optimization. However, as the overlap increases to 2 mm or more, preserving the shapes becomes increasingly challenging.

Digital Twin processing consistently improves contrast across all frame counts (Figure \ref{result_twopoint}). However, it does not significantly enhance resolution, indicating that sound pressure optimization plays a more crucial role in achieving high resolution than optical enhancements alone.

For overlaps of 2 mm or more, shape maintenance becomes difficult across all frame counts. Nevertheless, using nine or more frames results in relatively better shape preservation compared to cases with three or fewer frames. This indicates that higher frame counts can contribute to enhanced shape preservation under certain conditions. In conclusion, these results underscore the importance of optimizing sound pressure distribution to achieve high-resolution caustic patterns.

\subsection*{Noise Characteristics}
Our experiments reveal two main categories of noise that reduce the clarity of the generated caustic patterns, as illustrated in Figures~\ref{mechanism}(c), \ref{result_experiment}, and \ref{result_twopoint}. The first type consists of dark, spot-like artifacts that manifest in regions intended to be uniformly illuminated. These artifacts commonly arise from acoustic grating lobes inherent in the beamforming process of the transducer array, as well as from interference patterns caused by partial reflections at the tank walls or mesh screen holders. Such unintended acoustic pressure maxima introduce localized surface deformations in the liquid, leading to ring-shaped or blotchy dark regions on the caustic image. Although increasing the number of frames for time-averaging can partially mitigate these effects, small residual patches tend to remain, particularly in areas of high curvature or abrupt transitions in the caustic pattern.

A second form of noise appears as bright speckles in areas that should remain dark. These speckles often originate from subtle ripples on the liquid surface—caused by microscopic fluid motion—that focus or scatter the parallel light rays in unpredictable ways. In addition, minor misalignments in the Fresnel lens, mesh screen, or camera calibration can amplify such bright points, since even a small deviation in the optical path can redirect light into otherwise shadowed regions.

Figures~\ref{result_experiment} and \ref{result_twopoint} demonstrate that raising the frame count for time-averaging substantially alleviates both dark patches and bright speckles, resulting in improved global contrast. Nevertheless, there is an inherent trade-off between noise suppression and sharpness, as excessive averaging can soften the edges of the projected pattern, reducing the sharpness of the final caustic image. Taken together, these observations indicate that dark noise typically stems from unwanted grating lobes and interference, whereas bright speckles are more closely tied to local surface ripples and optical scattering. While time-averaging and Digital Twin optimization partly address both issues, neither approach fully resolves them, underscoring the need for more refined fluid modeling, transducer design, and optical alignment.

\subsection*{Limitations and Future Directions}
The results presented in this study highlight several inherent challenges in generating high-quality caustic images using ultrasonically modulated liquid surfaces (Figures \ref{animation}, \ref{Result_Focus_and_shadow}). Micro-scale disturbances and subtle ripples in the fluid—arising from factors such as surface tension, viscosity variations, and incomplete acoustic control—introduce unintended refractions and bright streaks in darker regions (Figures \ref{animation}, \ref{Result_Focus_and_shadow}). These irregularities hinder the formation of the sharp, high-contrast edges achievable with precisely machined solid optics, resulting in more diffuse patterns.

The fidelity of the generated images is also constrained by fundamental parameters, including transducer array spacing, operating frequency, and the size of the fluid surface. Currently, the optical outcomes cannot match the sharpness of static glass-based methods, leading to inherently softer and less defined patterns (Figures \ref{animation}, \ref{Result_Focus_and_shadow}). Scaling up the phased array configuration and the liquid surface dimensions is a promising direction to improve acoustic field control and potentially achieve finer pattern details.

Future work will focus on refining the loss function and optimization strategies to enable more nuanced grayscale transitions. Exploring techniques such as advanced beamforming, the use of metasurfaces, or adaptive algorithms could better balance large-scale shape control with the pursuit of finer detail. These advancements, while potentially increasing complexity and cost, represent steps towards novel forms of optical modulation through dynamic liquid interfaces.

Despite these challenges, this work demonstrates a new category of optically modulated surfaces and real-time computational optimization methods. With continued scaling, improved fluid–acoustic interaction models, and rigorous technical evaluation, this technology holds potential for applications in ambient displays, interactive installations, and novel optical deformation-based fabrication tools. This research lays a foundation for future innovations in dynamic caustic generation.

\section*{Methods}

\subsection*{Equipment Setup}
The experimental setup, as depicted in Figure \ref{explanation} (a) and detailed in Table \ref{table:setup}, was placed in a dark room to minimize ambient light interference. This vertically aligned system facilitated the simultaneous generation of optical and acoustic fields. A Phased Array Transducer (PAT) with 256 transducers (40 kHz, TAMURA, H2 50137) arranged in a 16 $\times$ 16 matrix was positioned 200 mm above the liquid medium's surface. The PAT was controlled by an FPGA board (Waveshare CoreEP4CE6) as described by Morales et al. \cite{Morales2021-wy}. The liquid medium consisted of 1 kg of transparent silicone oil (Shin-Etsu Sillicons, KF-96H-10000) contained within a transparent acrylic tank (155 mm length, 155 mm width, and 45 mm height). A point LED light source (ELEKIT LK-3WH) was placed below the tank and collimated using a Fresnel lens (Nihon Tokushu Kogaku Jushi, AH0498891). A nylon mesh screen (106 micrometers thick, 150 $\times$ 150 grid with 61 micrometers per grid) was positioned between the silicone oil tank and the PAT.

\subsection*{Image Acquisition and Camera Calibration}
Caustic patterns were captured using two cameras equipped with a PoC sensor kit (Sony Semiconductor Solutions Corporation) and FUJIFILM CF12ZA-1S lenses (F1.8/12 mm). The cameras were positioned diagonally to the projection screen to avoid obstruction by the Fresnel lens. These cameras operated at a resolution of 5320 $\times$ 4600 pixels and captured images at 6 frames per second with a gain of 30.0 dB and an exposure time of 6 ms.

To correct for perspective distortion, a camera calibration was performed using a standard checkerboard pattern and the OpenCV library. A series of checkerboard images at various orientations were captured. Four corresponding points were manually selected on the checkerboard in the captured image and a reference frontal view. These points were used to calculate a perspective transformation matrix using the \texttt{cv2.getPerspectiveTransform()} function. This matrix was subsequently applied to the captured caustic images using the \texttt{cv2.warpPerspective()} function to obtain a frontal view of the caustic patterns, ensuring accurate alignment and measurement. The transformation matrix was saved for later use.

\subsection*{Acoustic Hologram Generation and Upload}
The calculated phase delays were discretized and uploaded to the FPGA board controlling the PAT. Communication with the FPGA was established using the PySerial library. The phase values were sent to the respective transducers to generate the desired acoustic pressure field.

\subsection*{Dynamic Caustic Generation Procedure}
The software and packages used in this manuscript is as summarized in Table 2, and the procedure for generating dynamic caustic patterns are as follows:

\begin{enumerate}
    \item Target Image Preparation: A target grayscale image was loaded using the Pillow (PIL) library and converted to a NumPy array. The pixel values were inverted such that black areas corresponded to high acoustic pressure regions. The image was then resized to 192x192 pixels using bicubic interpolation via TensorFlow.
    \item Initial Phase Calculation: Initial phase delays for each transducer were calculated numerically using the Diff-PAT method \cite{acoustic_fushimi_2021}, as described in the Introduction. This involved optimizing the phase delays to produce the desired acoustic pressure distribution corresponding to the inverted target image. The TensorFlow library was used for the numerical optimization, employing the Adam optimizer \cite{kingma2014adam}.
    \item Liquid Surface Modulation: The generated acoustic pressure field deformed the surface of the silicone oil. The concave deformations acted as dynamic lenses, refracting the collimated light passing through the liquid.
    \item Caustic Pattern Capture: The resulting caustic pattern projected onto the mesh screen was captured by the cameras.
    \item Digital Twin Optimization: The captured image was processed using a Digital Twin framework implemented in TensorFlow. This involved applying the pre-calculated perspective transformation matrix and subtracting a background image to isolate the caustic pattern. The processed image was then compared to the target image using a cosine similarity loss function. The gradients of the loss function were calculated using automatic differentiation in TensorFlow, and the transducer phase delays were iteratively adjusted to minimize the loss, refining the caustic pattern.
    \item Time Averaging (Optional): For some experiments, the phase delays were updated sequentially over multiple frames (3, 9, or 24 frames) to implement a time-averaging technique, as described in the Results and Discussion section.
\end{enumerate}

The numerical optimization of the acoustic hologram is detailed in the Introduction. The Digital Twin optimization process, including the loss function and update rule, is also described in the Introduction. This section focuses on the experimental implementation of these methods.

\begin{table}[ht]
\centering
\caption{Summary of Software and Libraries}
\label{table:software}
\begin{tabularx}{\textwidth}{p{0.3\textwidth} X}
\toprule
\textbf{Software/Library} & \textbf{Description} \\
\midrule
OpenCV & Used for camera calibration, image processing, and perspective transformation. \\
TensorFlow & Used for numerical optimization of acoustic holograms and implementation of the Digital Twin framework. \\
NumPy & Used for numerical computations and array manipulation. \\
Pillow (PIL) & Used for image loading and manipulation. \\
PySerial & Used for serial communication with the FPGA board controlling the PAT. \\
SciPy & Used for loading transducer array data. \\
Matplotlib & Used for generating plots. \\
\bottomrule
\end{tabularx}
\end{table}

\begin{table}[ht]
\centering
\caption{Summary of the Experimental Setup}
\label{table:setup}
\begin{tabularx}{\textwidth}{p{0.25\textwidth} X}
\toprule
\textbf{Component} & \textbf{Specification / Description} \\
\midrule
\vspace{3pt} \\
\textbf{Phased Array Transducer}
\vspace{2pt} &
256 transducers (40\,kHz, TAMURA, H2\,50137) arranged in a 16\,$\times$\,16 matrix;
positioned 200\,mm above the liquid surface;
controlled by FPGA (Waveshare CoreEP4CE6)\cite{Morales2021-wy}.
\vspace{3pt} \\

\textbf{Liquid Medium}
\vspace{2pt} &
1\,kg transparent silicone oil (Shin-Etsu KF-96H-10000) in an acrylic tank
(155\,mm\,$\times$\,155\,mm\,$\times$\,45\,mm);
chosen for viscosity and optical clarity.
\vspace{3pt} \\

\textbf{Light Source}
\vspace{2pt} &
Point LED (ELEKIT LK-3WH) placed below the tank.
\vspace{3pt} \\

\textbf{Fresnel Lens}
\vspace{2pt} &
Nihon Tokushu Kogaku Jushi (AH0498891), used to collimate LED light
into a parallel beam.
\vspace{3pt} \\

\textbf{Projection Screen}
\vspace{2pt} &
Nylon mesh screen (106\,$\mu$m thick, grid: 150\,$\times$\,150 with 61\,$\mu$m per grid)
between tank and PAT; transmits acoustic waves and supports light projection.
\vspace{3pt} \\

\textbf{Cameras}
\vspace{2pt} &
Two cameras with PoC sensor kit (Sony), FUJIFILM CF12ZA-1S lens (F1.8/12\,mm);
resolution: 5320\,$\times$\,4600, 6\,fps, gain: 30.0\,dB, exposure: 6\,ms;
positioned diagonally to avoid Fresnel lens interference.
\vspace{3pt} \\

\bottomrule
\end{tabularx}
\end{table}

\newpage

\bibliography{main}

\begin{thebibliography}{10}
\urlstyle{rm}
\expandafter\ifx\csname url\endcsname\relax
  \def\url#1{\texttt{#1}}\fi
\expandafter\ifx\csname urlprefix\endcsname\relax\def\urlprefix{URL }\fi
\expandafter\ifx\csname doiprefix\endcsname\relax\def\doiprefix{DOI: }\fi
\providecommand{\bibinfo}[2]{#2}
\providecommand{\eprint}[2][]{\url{#2}}

\bibitem{Weinstein1969-de}
\bibinfo{author}{Weinstein, L.~A.} \& \bibinfo{author}{Beckmann, P.}
\newblock \emph{\bibinfo{title}{Open resonators and open waveguides}} (\bibinfo{publisher}{The Golem Press}, \bibinfo{address}{Boulder (Colo.)}, \bibinfo{year}{1969}).

\bibitem{Yue2012-nn}
\bibinfo{author}{Yue, Y.}, \bibinfo{author}{Iwasaki, K.}, \bibinfo{author}{Chen, B.-Y.}, \bibinfo{author}{Dobashi, Y.} \& \bibinfo{author}{Nishita, T.}
\newblock \bibinfo{journal}{\bibinfo{title}{Pixel art with refracted light by rearrangeable sticks}}.
\newblock {\emph{\JournalTitle{Comput. Graph. Forum}}} \textbf{\bibinfo{volume}{31}}, \bibinfo{pages}{575--582} (\bibinfo{year}{2012}).

\bibitem{Yue2014-hs}
\bibinfo{author}{Yue, Y.}, \bibinfo{author}{Iwasaki, K.}, \bibinfo{author}{Chen, B.-Y.}, \bibinfo{author}{Dobashi, Y.} \& \bibinfo{author}{Nishita, T.}
\newblock \bibinfo{journal}{\bibinfo{title}{{Poisson-Based} continuous surface generation for {Goal-Based} caustics}}.
\newblock {\emph{\JournalTitle{ACM Trans. Graph.}}} \textbf{\bibinfo{volume}{33}}, \bibinfo{pages}{1--7} (\bibinfo{year}{2014}).

\bibitem{Schwartzburg2014-oy}
\bibinfo{author}{Schwartzburg, Y.}, \bibinfo{author}{Testuz, R.}, \bibinfo{author}{Tagliasacchi, A.} \& \bibinfo{author}{Pauly, M.}
\newblock \bibinfo{journal}{\bibinfo{title}{High-contrast computational caustic design}}.
\newblock {\emph{\JournalTitle{ACM Trans. Graph.}}} \textbf{\bibinfo{volume}{33}}, \bibinfo{pages}{1--11} (\bibinfo{year}{2014}).

\bibitem{Suzuki2019-xj}
\bibinfo{author}{{Suzuki}}, \bibinfo{author}{{Fujisawa}} \& \bibinfo{author}{{Mikawa}}.
\newblock \bibinfo{title}{Simulation controlling method for generating desired water caustics}.
\newblock In \emph{\bibinfo{booktitle}{2019 International Conference on Cyberworlds ({CW})}}, vol.~\bibinfo{volume}{0}, \bibinfo{pages}{163--170} (\bibinfo{year}{2019}).

\bibitem{Hoshi2010}
\bibinfo{author}{Hoshi, T.}, \bibinfo{author}{Takahashi, M.}, \bibinfo{author}{Iwamoto, T.} \& \bibinfo{author}{Shinoda, H.}
\newblock \bibinfo{journal}{\bibinfo{title}{Noncontact tactile display based on radiation pressure of airborne ultrasound}}.
\newblock {\emph{\JournalTitle{IEEE Transactions on Haptics}}} \textbf{\bibinfo{volume}{3}}, \bibinfo{pages}{155--165}, \doiprefix\url{10.1109/TOH.2010.4} (\bibinfo{year}{2010}).

\bibitem{Long2014}
\bibinfo{author}{Long, B.}, \bibinfo{author}{Seah, S.~A.}, \bibinfo{author}{Carter, T.} \& \bibinfo{author}{Subramanian, S.}
\newblock \bibinfo{journal}{\bibinfo{title}{Rendering volumetric haptic shapes in mid-air using ultrasound}}.
\newblock {\emph{\JournalTitle{ACM Transactions on Graphics}}} \textbf{\bibinfo{volume}{33}}, \bibinfo{pages}{1--10}, \doiprefix\url{10.1145/2661229.2661257} (\bibinfo{year}{2014}).

\bibitem{monnai2014haptomime}
\bibinfo{author}{Monnai, Y.} \emph{et~al.}
\newblock \bibinfo{title}{Haptomime: mid-air haptic interaction with a floating virtual screen}.
\newblock In \emph{\bibinfo{booktitle}{Proceedings of the 27th annual ACM symposium on User interface software and technology}}, \bibinfo{pages}{663--667} (\bibinfo{year}{2014}).

\bibitem{Hirayama2019}
\bibinfo{author}{Hirayama, R.}, \bibinfo{author}{Plasencia, D.~M.}, \bibinfo{author}{Masuda, N.} \& \bibinfo{author}{Subramanian, S.}
\newblock \bibinfo{journal}{\bibinfo{title}{A volumetric display for visual, tactile and audio presentation using acoustic trapping}}.
\newblock {\emph{\JournalTitle{Nature}}} \textbf{\bibinfo{volume}{575}}, \bibinfo{pages}{320--323}, \doiprefix\url{10.1038/s41586-019-1739-5} (\bibinfo{year}{2019}).

\bibitem{Fushimi2019a}
\bibinfo{author}{Fushimi, T.}, \bibinfo{author}{Marzo, A.}, \bibinfo{author}{Drinkwater, B.~W.} \& \bibinfo{author}{Hill, T.~L.}
\newblock \bibinfo{journal}{\bibinfo{title}{Acoustophoretic volumetric displays using a fast-moving levitated particle}}.
\newblock {\emph{\JournalTitle{Applied Physics Letters}}} \textbf{\bibinfo{volume}{115}}, \bibinfo{pages}{064101}, \doiprefix\url{10.1063/1.5113467} (\bibinfo{year}{2019}).

\bibitem{Koroyasu2023-ws}
\bibinfo{author}{Koroyasu, Y.} \emph{et~al.}
\newblock \bibinfo{journal}{\bibinfo{title}{Microfluidic platform using focused ultrasound passing through hydrophobic meshes with jump availability}}.
\newblock {\emph{\JournalTitle{PNAS Nexus}}} \textbf{\bibinfo{volume}{2}}, \bibinfo{pages}{gad207} (\bibinfo{year}{2023}).

\bibitem{Morales2021-wy}
\bibinfo{author}{Morales, R.}, \bibinfo{author}{Ezcurdia, I.}, \bibinfo{author}{Irisarri, J.}, \bibinfo{author}{Andrade, M. A.~B.} \& \bibinfo{author}{Marzo, A.}
\newblock \bibinfo{journal}{\bibinfo{title}{Generating airborne ultrasonic amplitude patterns using an open hardware phased array}}.
\newblock {\emph{\JournalTitle{NATO Adv. Sci. Inst. Ser. E Appl. Sci.}}} \textbf{\bibinfo{volume}{11}}, \bibinfo{pages}{2981} (\bibinfo{year}{2021}).

\bibitem{acoustic_fushimi_2021}
\bibinfo{author}{Fushimi, T.}, \bibinfo{author}{Yamamoto, K.} \& \bibinfo{author}{Ochiai, Y.}
\newblock \bibinfo{journal}{\bibinfo{title}{Acoustic hologram optimisation using automatic differentiation.}}
\newblock {\emph{\JournalTitle{Scientific Reports}}} \doiprefix\url{10.1038/S41598-021-91880-2} (\bibinfo{year}{2021}).

\bibitem{Fushimi2024-ih}
\bibinfo{author}{Fushimi, T.}, \bibinfo{author}{Tagami, D.}, \bibinfo{author}{Yamamoto, K.} \& \bibinfo{author}{Ochiai, Y.}
\newblock \bibinfo{journal}{\bibinfo{title}{A digital twin approach for experimental acoustic hologram optimization}}.
\newblock {\emph{\JournalTitle{Communications Engineering}}} \textbf{\bibinfo{volume}{3}}, \bibinfo{pages}{1--8} (\bibinfo{year}{2024}).

\bibitem{kingma2014adam}
\bibinfo{author}{Kingma, D.~P.} \& \bibinfo{author}{Ba, J.~L.}
\newblock \bibinfo{journal}{\bibinfo{title}{Adam: A method for stochastic optimization}}.
\newblock {\emph{\JournalTitle{3rd International Conference on Learning Representations, ICLR 2015 - Conference Track Proceedings}}} \textbf{\bibinfo{volume}{1}}, \bibinfo{pages}{1--15} (\bibinfo{year}{2015}).

\bibitem{elizondo2023}
\bibinfo{author}{Elizondo, S.}, \bibinfo{author}{Ezcurdia, I.}, \bibinfo{author}{Goñi, J.}, \bibinfo{author}{Galar, M.} \& \bibinfo{author}{Marzo, A.}
\newblock \bibinfo{journal}{\bibinfo{title}{{Enhancing the quality of amplitude patterns using time-multiplexed virtual acoustic fields}}}.
\newblock {\emph{\JournalTitle{Applied Physics Letters}}} \textbf{\bibinfo{volume}{123}}, \bibinfo{pages}{154102}, \doiprefix\url{10.1063/5.0164657} (\bibinfo{year}{2023}).
\newblock \eprint{https://pubs.aip.org/aip/apl/article-pdf/doi/10.1063/5.0164657/18166719/154102\_1\_5.0164657.pdf}.

\bibitem{Plasencia2020}
\bibinfo{author}{Plasencia, D.~M.}, \bibinfo{author}{Hirayama, R.}, \bibinfo{author}{Montano-Murillo, R.} \& \bibinfo{author}{Subramanian, S.}
\newblock \bibinfo{journal}{\bibinfo{title}{Gs-pat: High-speed multi-point sound-fields for phased arrays of transducers}}.
\newblock {\emph{\JournalTitle{ACM Trans. Graph.}}} \textbf{\bibinfo{volume}{39}}, \doiprefix\url{0.1145/3386569.3392492} (\bibinfo{year}{2020}).

\bibitem{deep_lee_2022}
\bibinfo{author}{Lee, M.}, \bibinfo{author}{Lew, H.~M.}, \bibinfo{author}{Youn, S.}, \bibinfo{author}{Kim, T.} \& \bibinfo{author}{Hwang, J.~Y.}
\newblock \bibinfo{journal}{\bibinfo{title}{Deep learning-based framework for fast and accurate acoustic hologram generation}}.
\newblock {\emph{\JournalTitle{IEEE Transactions on Ultrasonics, Ferroelectrics and Frequency Control}}} \doiprefix\url{10.1109/TUFFC.2022.3219401} (\bibinfo{year}{2022}).

\bibitem{Peli1990-ah}
\bibinfo{author}{Peli, E.}
\newblock \bibinfo{journal}{\bibinfo{title}{Contrast in complex images}}.
\newblock {\emph{\JournalTitle{J. Opt. Soc. Am. A}}} \textbf{\bibinfo{volume}{7}}, \bibinfo{pages}{2032--2040} (\bibinfo{year}{1990}).

\end{thebibliography}



\section*{Acknowledgements}

We would like to extend our sincere gratitude to Sony Semiconductor Solutions Corporation for their generous support in lending us a sensor kit, which was instrumental in the completion of our research. Additionally, we deeply appreciate the efforts of our lab member, Mr. Takumi Yokoyama, for his exceptional work in capturing the equipment and results beautifully. We gratefully acknowledge the support of AI tools, OpenAI’s GPT-4, and Anthropic’s Claude. The authors have diligently reviewed and verified all generated outputs to ensure their accuracy and relevance.

\section*{Data availability}
All data generated or analyzed during this study, including experimental measurements, simulation results, and analysis scripts, are available in the Digital Nature Group repository at \url{https://github.com/DigitalNatureGroup/Dynamic-Caustics-by-Ultrasonically-Modulated-Liquid-Surface}. 

\section*{Author contributions statement}
K.N., T.F., and Y.O. conceived the research. K.N. designed the experiments, conducted the experimental work, and acquired the data. K.N. and A.T. contributed to data analysis. T.F. and Y.O. supervised the project and contributed to the interpretation of the results. All authors contributed to writing, revising, and final approval of the manuscript.

\end{document}